\documentclass[twocolumn,showpacs,prb,amsmath,amssymb,longbibliography]{revtex4-2}
\usepackage{graphicx}
\usepackage[utf8]{inputenc}
\usepackage{braket}       
\usepackage{amsmath}

\begin{document}

\title{Internal dynamics and dielectric screening of confined multiexciton states}

\author{Josep Planelles}
\affiliation{Dept. de Qu\'imica F\'isica i Anal\'itica, Universitat Jaume I, 12080, Castell\'o, Spain}

\author{Juan I. Climente}
\affiliation{Dept. de Qu\'imica F\'isica i Anal\'itica, Universitat Jaume I, 12080, Castell\'o, Spain}

\author{Jos\'e L. Movilla}
\affiliation{Dept. d'Educaci\'o i Did\`actiques Espec\'ifiques, Universitat Jaume I, 12080, Castell\'o, Spain}
\email{movilla@uji.es}

\date{\today}

\begin{abstract}
Recent experimental and computational studies suggest that biexcitons (BX) confined in large CsPbBr$_3$ nanocrystals 
experience reduced dielectric screening as compared to excitons (X) and trions (X$^*$). Here we provide a physical rationale
to explain such a behavior. A characteristic frequency is introduced, which describes the internal dynamics of an exciton 
within the excitonic complex. By means of effective mass--variational Quantum Monte Carlo simulations, 
we show that, in large nanocrystals, the frequency is similar for X and X$^*$, but smaller for BX. 
Because the frequencies exceed that of the bulk longitudinal optical phonon, this leads to a reduced dielectric constant
for BX, which is in contrast with the behavior of strongly confined nanocrystals.
\end{abstract}

\maketitle

\section{Introduction}

Metal halide perovskite nanocrystals (NCs) have emerged as promising materials for optoelectronic applications due to their bright, 
color-pure and tunable emission, which combines with facile material synthesis.\cite{DeyACS,SuhailJPCM,LiuAM,DongEL}
Understanding the energetics and dynamics of multiexciton states in these nanostructures is a prerequesite 
to fully exploit their potential, since they are expected to be populated under device relevant conditions.\cite{ShulenbergerJPCL}
Furthermore, certain materials such as CsPbBr$_3$, have witnessed the formation of ensembles of large 
--yet highly emissive-- NCs, with dimensions exceeding the Bohr diameter ($\sim 7$ nm). 
Unlike conventional colloidal quantum dots, these NCs are in the weak confinement regime, 
which brings about a distinct physical response\cite{ButkusCM,ZhuNAT}. 
An important consequence is the suppression of non-radiative Auger processes,\cite{UtzatSCI,HuangEL} 
which makes charged and multiexciton states have a more active participation in the radiative properties.

Motivated by the aforementioned situation, in the last years several studies have investigated the nature
of charged excitons (trions, X$^*$) and biexcitons (BX) in large metal halide perovskite NCs.\cite{TamaratNC,FuNL,BeckerNAT,WangAM,MakarovNL,CastanedaACS,AneeshJPCc,YumotoJPCL,AshnerACSener,HuangJPCL,ShenJPCc,DanaACS,PooniaPRB,LubinACS,AmaraNL,ZhuAM,ChoACS,KazesNL,NguyenPRB,StrandellJCP,BarfusserNL,BarfusserACSph,ParkPRM,MovillaPRB}
Significant attention has been paid to the study of binding energies in these excitonic complexes.
Photoluminescence spectroscopy of individual CsPbBr$_3$ NCs with lateral sizes in the range of $7-20$ nm,
measured at cryogenic temperatures, revealed trion binding energies ($\Delta_{X^*}$) of $7-25$ meV,
and biexciton binding energies ($\Delta_{BX}$) of $25-40$ meV.\cite{TamaratNC,FuNL,BeckerNAT,ZhuAM,ChoACS,AmaraNL}
The latter are surprisingly large for weakly confined nanostructures.
Effective mass models including electronic correlations, which successfully describe the 
confinement dependence of exciton ($\Delta_X$) and trion ($\Delta_{X^*}$) binding energies in these systems,
underestimate $\Delta_{BX}$ by almost a factor of 2.\cite{ZhuAM,ChoACS,NguyenPRB,ParkPRM,MovillaPRB}\\

To account for the discrepancy between experiments and theoretical models, Refs.~\cite{ZhuAM} and \cite{ChoACS}
\emph{postulated} that BX experience a weaker dielectric screening than X. 
A similar conclusion is reached when the polaronic nature of interactions is taken into account.
Within Haken theory\cite{Haken,BaranowskiAEM}, electrons and holes polarize the ionic lattice efficiently beyond
the so-called polaron radius (yielding a static dielectric constant) but less so under it (yielding a high-frequency dielectric constant).
Considering this distance dependence, Ref.~\cite{MovillaPRB} observed that the longer 
electron-hole distances in BX --as compared to X-- lead to small $\Delta_{BX}$ values. 
The conclusion was that BX should have larger polaron radius than X, in order to reproduce the 
experimentally observed $\Delta_{BX}$ values.
In all cases, a physical rationale for the different dielectric constant (or polaron radius) 
of X and BX is missing.
Recent spectroscopic and DFT studies suggest that photoexcitations gives rise to phonon-mediated 
attractive exciton-exciton interactions, but this phenomenon is weak in CsPbBr$_3$,\cite{YazdaniNP}
so the question remains open.\\

To address this problem, in the present work we introduce the concept of an \emph{effective exciton} 
within a multiexciton system, defined as a reduced two-body subsystem (one electron and one hole) 
embedded in a larger interacting quantum state. The effective exciton is described by 
the reduced density matrix, by tracing out the degrees of freedom of the remaining particles. 
The intrinsic dynamical behavior of this effective exciton can be characterized by a 
\emph{characteristic frequency} associated with a suitably chosen observable $ \hat{A} $, defined via:
\begin{equation}
	\omega_A = \frac{1}{\hbar} \frac{ \sqrt{ \langle [\hat{H}, \hat{A}]^2 \rangle}}{ \sqrt{ \langle \hat{A}^2 \rangle }}.
\end{equation}

By setting $\hat{A}=r_{eh}$, with $r_{eh}$ the electron-hole distance of the effective exciton, 
this formalism allows us to estimate the internal timescale of exciton dynamics in the presence of additional charges. 
Using variational quantum Monte Carlo wavefunctions, we compute this characteristic frequency for confined X, X$^*$ and BX in CsPbBr$_3$ NCs. 
When applied to weakly confined NCs, the model reveals that $\omega_X \sim \omega_{X^*}$, but $\omega_{BX} < \omega_X$, 
which evidences that excitons within BX experience a slower effective internal motion.\\

 This result naturally connects with the known electric field frequency dependence of the dielectric response in polar semiconductors.\cite{YuCardona,Fox2010} 
 The dielectric function $ \varepsilon(\omega) = \varepsilon_R (\omega) + i\, \varepsilon_I (\omega) $ is a complex quantity whose real part 
 ($\varepsilon_R$) governs the electrostatic screening.
 %
 Internal electron-hole motion with frequencies below and above those of the relevant lattice polarization mechanisms lead to different screening.
 In large CsPbBr$_3$ NCs, we find that $\omega_{X}$, $\omega_{X^*}$ and $\omega_{BX}$ exceed the bulk longitudinal optical phonon mode, $\omega_{LO}$. 
 Under these circumstances, a damped Lorenz oscillator function for $\varepsilon(\omega)$ shows that $\varepsilon_R^{BX} < \varepsilon_R^X \approx \varepsilon_R^{X^*}$.

\section{Theoretical Model}
\label{s:theo}

Excitonic complexes exert a fluctuating electric field on the lattice, 
whose frequency determines the dielectric screening.
Our goal in this section is to derive expressions for such frequencies.
The electric field is critically related to the internal motion of electron-hole pairs,
which is in turn influenced by the correlations among all the electrons and holes in the complex.
We can however simplify this many-body problem by drawing a parallel with Bader's theory of atoms in molecules,\cite{Bader_book,KumarJCS} 
and define an exciton as part of an excitonic complex (e.g. trion or biexciton) in a given medium. 
 We thus define the \emph{effective exciton} as an entity described by a density matrix in which 
we have integrated over all coordinates except those of one electron and one hole. 
 And this can be done even though it is not isolated, just as atoms in molecules are not. 
We can then calculate, for example, its characteristic frequency. 
This calculation is equivalent to the calculation of the characteristic frequency 
of one of the electrons and one of the holes of the system using the total wavefunction, 
which for simplicity is the approach we follow here.

\subsection{Oscillation frequency of an observable}
Let us consider a quantum mechanical time-independent Hermitian operator $\hat{A}$ and a quantum system in a stationary state $\ket{\psi_0}$. We are interested in quantifying the characteristic frequency associated with the quantum dynamics of $\hat{A}$, even though its expectation value is constant in time. To this end, we introduce the following definition of a squared frequency associated with the observable $\hat{A}$:
\begin{equation}
\label{Mod1}
\omega_A^2 := \frac{\langle [\hat{H}, \hat{A}]^2 \rangle}{\hbar^2 \langle \hat{A}^2 \rangle}.
\end{equation}

We should stress that although in a stationary state $\ket{\psi_0}$ the expectation value of a time-independent operator remains constant, this does not imply that the observable itself is static: internal quantum fluctuations persist, and these manifest in non-trivial variances and quantum uncertainties. Thus, although $\langle \hat{A} \rangle$ is time-independent, quantities like $\langle \hat{A}^2 \rangle$, the commutator $[\hat{H}, \hat{A}]$, and nested commutators such as $[\hat{H}, [\hat{H}, \hat{A}]]$ encode dynamical information.\\

In analogy with classical harmonic motion, where the acceleration is proportional to the displacement (i.e., $\ddot{x} = -\omega^2 x$), a quantum observable $\hat{A}$ can exhibit similar behavior when one analyzes the spectral spread of the transitions it induces, as encoded in its variance and transition matrix elements. Indeed, the matrix elements $\langle \psi_n | \hat{A} | \psi_0 \rangle$  oscillate at frequencies $\omega_{n0} = \frac{E_n - E_0}{\hbar}$.\\

The variance of $\hat{A}$ in the stationary state $\ket{\psi_0}$ reads: ${\rm Var}(\hat{A}) = \langle \hat{A}^2 \rangle - \langle \hat{A} \rangle^2$. Since $\langle \hat{A} \rangle$ is constant in time, the oscillatory or dynamical behavior must be contained in $\langle \hat{A}^2 \rangle$ and higher moments.\\

Following the harmonic motion classical analogy, we calculate $\frac{d^2 \hat{A}}{dt^2} $ using the Heisenberg equation of motion for the time-independent operator $\hat{A}$:
\begin{equation}
\label{Mod2}
\frac{d\hat{A}}{dt} = \frac{i}{\hbar} [\hat{H}, \hat{A}] ,
\end{equation}

Applying this twice and then taking expectation values in $\ket{\psi_0}$, we obtain:
\begin{equation}
\label{Mod3}
    \left\langle \frac{d^2}{dt^2} \hat{A}(t) \right\rangle = -\frac{1}{\hbar^2} \langle [\hat{H}, [\hat{H}, \hat{A}]] \rangle.
\end{equation}

Finally, inserting $ \mathbb{I} = \sum_n \ket{\psi_n}\bra{\psi_n}$ --the completeness relation--  into the double commutator, we end up with:
\begin{equation}
\label{Mod4}
    \langle \psi_0 | [\hat{H}, [\hat{H}, \hat{A}]] | \psi_0 \rangle = \sum_n (E_n - E_0)^2 |\langle \psi_n | \hat{A} | \psi_0 \rangle|^2.
\end{equation}

Similarly, we find:
\begin{equation}
\label{Mod5}
    \langle \psi_0 | \hat{A}^2 | \psi_0 \rangle = \sum_n |\langle \psi_n | \hat{A} | \psi_0 \rangle|^2.
\end{equation}

Then, although $[\hat{H}, [\hat{H}, \hat{A}]] \neq [\hat{H}, \hat{A}]^2$, their expectation values in a stationary state satisfy $ \langle \psi_0 | [\hat{H}, \hat{A}]^2| \psi_0 \rangle=\sum_n |\langle \psi_0 | [\hat{H}, \hat{A}]|\psi_n \rangle|^2=\langle \psi_0 | [\hat{H}, [\hat{H}, \hat{A}]] | \psi_0 \rangle $. Therefore, the ratio
\begin{equation}
\label{Mod6}
    \frac{\langle [\hat{H}, \hat{A}]^2 \rangle}{\hbar^2 \langle \hat{A}^2 \rangle} = \frac{\sum_n (E_n - E_0)^2 |\langle \psi_n | \hat{A} | \psi_0 \rangle|^2}{\hbar^2 \sum_n |\langle \psi_n | \hat{A} | \psi_0 \rangle|^2}
\end{equation}

\noindent provides a natural frequency scale characterizing the quantum observable \(\hat{A}\) in the stationary state $\ket{\psi_0}$. This expression can be interpreted as a weighted average of the squared transition frequencies $\omega_{n0} = (E_n - E_0)/\hbar$, where:
\begin{equation}
\label{Mod7}
\omega_A^2 = \sum_n w_n \, \omega_{n0}^2, \quad \text{with} \quad w_n = \frac{|\langle \psi_n | \hat{A} | \psi_0 \rangle|^2}{\sum_m |\langle \psi_m | \hat{A} | \psi_0 \rangle|^2}.
\end{equation}

Physically, this means that although the observable $\hat{A}$ in the stationary state $\ket{\psi_0}$ does not oscillate at a single frequency, its quantum fluctuations involve a superposition of oscillations at multiple frequencies $\omega_{n0}$. Thus, $\omega_A$ provides a \textit{characteristic} or \textit{effective} frequency scale capturing the overall spectrum of these fluctuations. Unlike a classical harmonic oscillator with a single oscillation frequency, the quantum observable \(\hat{A}(t)\) in a stationary state exhibits a weighted average over a \emph{spectrum} of transition frequencies.

\subsection{Characteristic frequency of X, X$^*$ and BX}

Let us consider a confined excitonic complex with electrons $1,2,\dots$ and holes $a,b,\ldots$,
defined by a state $\Psi(\mathbf{X})$. 
\(\mathbf{X}\) represents all the coordinates of the system: 
\( (\mathbf{r}_1, \mathbf{R}_a)\) for an exciton, 
\( (\mathbf{r}_1, \mathbf{r}_2, \mathbf{R}_a)\) for a negative trion, 
\( (\mathbf{r}_1, \mathbf{r}_2, \mathbf{R}_a, \mathbf{R}_b)\) for BX, etc. 
 In this section we aim at obtaining practical expressions for the frequency related to the 
internal motion between the electron $1$ and the hole $a$, which is our effective exciton,
under the influence of the other particles.\\

We consider an effective mass Hamiltonian $\hat{H}$ with real-valued eigenfunctions $\Psi(\mathbf{X})$.
This is often a good description of low-energy states in nanocrystals, 
even in the presence of strong electronic correlations.\cite{ChoACS,ZhuAM,NguyenPRB,MovillaPRB,RajadellPRB,MaciasNS,ClimenteJPCc,ClimenteJPCL,ChoPRL}
 The relevant observable is the distance between the chosen electron-hole pair, $\hat{A}=r_{1a}$. 
Building on Eq.~(\ref{Mod1}), the elements needed for the calculation of \( \omega_A \) 
are the denominator, $ \langle r_{1a}^2 \rangle $, and the numerator, $\langle [\hat{H}, r_{1a}]^2 \rangle$.
The denominator is simply given by: 
\begin{equation}
	\label{eq:den}
	\langle \hat{r}_{1a}^2 \rangle = \int d\mathbf{r}_1 \cdots d\mathbf{R}_b \, |\Psi(\mathrm{X})|^2 \, r_{1a}^2.
\end{equation}
Because the number of coordinates can be quite large for multiexciton systems, 
it is convenient to evaluate Eq.~(\ref{eq:den}) by defining the local observable:
\begin{equation}
\label{EqCharF9den}
	O_D(\Psi(\mathbf{X})) = r_{1a}^2.
\end{equation}
\noindent over the probability distribution $\Psi^2(\mathbf{X})$, 
and integrate it via a Metropolis algorithm.\\

As for the numerator,
\begin{equation}
\label{EqCharF1}
[\hat{H}, \hat{A}] = \frac{i\hbar}{m_e} \hat{\mathbf{p}}_1 - \frac{i\hbar}{m_h} \hat{\mathbf{p}}_a,
\end{equation}
\noindent with $m_e$ and $m_h$ the electron and hole masses. Then:
\begin{equation}
\label{EqCharF2}
\langle [\hat{H}, \hat{A}]^2 \rangle = \hbar^2 \left\langle \left( \frac{\hat{\mathbf{p}}_1}{m_e} - \frac{\hat{\mathbf{p}}_a}{m_h} \right)^2 \right\rangle.
\end{equation}
Given that $\Psi$ is real-valued, the expectation value can be written as:
\begin{equation}
\label{EqCharF3}
\left\langle \left( \frac{\hat{\mathbf{p}}_1}{m_e} - \frac{\hat{\mathbf{p}}_a}{m_h} \right)^2 \right\rangle
= -\hbar^2 \int d\mathbf{X} \, \Psi(\mathbf{X}) 
\left( \frac{\nabla_1}{m_e} - \frac{\nabla_a}{m_h} \right)^2 \Psi(\mathbf{X}).
\end{equation}

We can integrate by parts 
(with boundary terms vanishing, because $\Psi(\mathbf{X}) \to 0$ when $|\mathbf{X}| \to \infty$)
 to obtain:
\begin{equation}
\label{EqCharF4}
\int \Psi (\nabla \cdot \mathbf{G}) \, d\mathbf{X} = -\int (\nabla \Psi) \cdot \mathbf{G} \, d\mathbf{X},
\end{equation}
\noindent where $\mathbf{G} = \frac{\nabla_1 \Psi}{m_e} - \frac{\nabla_a \Psi}{m_h}$ and $\nabla = \left( \frac{\nabla_1}{m_e} - \frac{\nabla_a}{m_h} \right)$.
 In our case:
\begin{multline}
\label{EqCharF5}
-\hbar^2 \int d\mathbf{X} \, \Psi(\mathbf{X}) 
\left( \frac{\nabla_1}{m_e} - \frac{\nabla_a}{m_h} \right)^2 \Psi(\mathbf{X}) = \\
 = \hbar^2 \int \left| \left( \frac{\nabla_1}{m_e} - \frac{\nabla_a}{m_h} \right) \Psi \right|^2 d\mathbf{X}.
\end{multline}
%

A more convenient expression can be derived by replacing the derivatives as $\nabla_i \Psi = \Psi \nabla_i \ln \Psi$.
This yields:
\begin{equation}
\label{EqCharF6}
\left( \frac{\nabla_1}{m_e} - \frac{\nabla_a}{m_h} \right) \Psi 
= \Psi \left( \frac{\nabla_1 \ln \Psi}{m_e} - \frac{\nabla_a \ln \Psi}{m_h} \right),
\end{equation}
\noindent and the integrand becomes:
\begin{equation}
\label{EqCharF7}
\left| \left( \frac{\nabla_1}{m_e} - \frac{\nabla_a}{m_h} \right) \Psi \right|^2 
= \Psi^2 \left| \frac{\nabla_1 \ln \Psi}{m_e} - \frac{\nabla_a \ln \Psi}{m_h} \right|^2.
\end{equation}
\noindent Consequently:
\begin{multline}
\label{EqCharF8}
\left\langle \left( \frac{\hat{\mathbf{p}}_1}{m_e} - \frac{\hat{\mathbf{p}}_a}{m_h} \right)^2 \right\rangle 
= \\
	\hbar^2 \int d\mathbf{X} \, \Psi^2(\mathbf{X}) 
\left| \frac{1}{m_e} \nabla_1 \ln \Psi(\mathbf{X}) - \frac{1}{m_h} \nabla_a \ln \Psi(\mathbf{X}) \right|^2.
\end{multline}
We can say that this is the expectation value of the local observable:
\begin{equation}
\label{EqCharF9}
O_N(\Psi(\mathbf{X})) = \left| \frac{1}{m_e} \nabla_1 \ln \Psi(\mathbf{X}) - \frac{1}{m_h} \nabla_a \ln \Psi(\mathbf{X}) \right|^2,
\end{equation}
\noindent over the probability distribution $\Psi^2(\mathbf{X})$. 
It is then possible to integrate Eq.~(\ref{EqCharF8}) using a Metropolis algorithm.\\

In this work, we calculate the ground state of different excitonic complexes confined in cuboidal NCs 
by means of an effective mass - Variational Quantum Monte Carlo method. 
The method, which is described in detail in Refs.~\cite{PlanellesCPC,MovillaPRB}, uses Metropolis algorithm to compute local energies.
Here, the computation is extended to evaluate $O_N(\Psi(\mathbf{X}))$ and $O_D(\Psi(\mathbf{X}))$ as well,
and subsequently infer $\omega_A$ through Eq.~(\ref{Mod1}).
 In the following subsections we present the formulas to be implemented for $O_N$ in the case 
 of X, X$^*$ and BX systems  ($O_D$ is straightforward). 

\subsubsection{Confined exciton}

Consider an X formed by an electron at position \(\mathbf{r}_1$ 
and a hole at position \(\mathbf{R}_a\), 
 with wavefunction:\cite{MovillaPRB}
\begin{equation}
\label{EqCE1}
\Psi_X(\mathbf{r}_1, \mathbf{R}_a) = \psi_{\text{box}}(\mathbf{r}_1)\, \psi_{\text{box}}(\mathbf{R}_a)\, e^{-Z r_{1a}},
\end{equation}
\noindent Here, \(\psi_{\text{box}}(\mathbf{r}) = \cos(k_x x)\cos(k_y y)\cos(k_z z)\), with $k_\mu=2\pi/L_\mu$, are particle in the box functions
for non-interacting electron and holes in a cuboidal quantum dot with edge lengths ($L_x, L_y, L_z$).  
These provide a valid limit for $\Psi_X$ in the strongly confinement regime.
In turn, $e^{-Z r_{1a}}$ --with \(r_{1a} = |\mathbf{r}_1 - \mathbf{R}_a|\)-- is a hydrogen-like function, 
with variational parameter $Z$, which provides a valid limit in the weakly confinement regime.
 For the above function, $O_N$ is broken down as:
\begin{equation}
\label{EqCE2}
O_N(\Psi_X(\mathbf{X})) = \sum_{\mu = x,y,z} \left( \frac{1}{m_e} \frac{\partial \ln \Psi_X}{\partial r_{1\mu}} - \frac{1}{m_h} \frac{\partial \ln \Psi_X}{\partial R_{a\mu}} \right)^2.
\end{equation}
\noindent After some algebra, the expression becomes:\cite{SuppMat}
\begin{multline}
\label{EqCE3}
O_N(\Psi_X(\mathbf{X})) = 
	\sum_{\mu = x,y,z} \left[ - \frac{k_\mu}{m_e} \tan(k_\mu r_{1\mu})  + \right. \\
	\left.  + \frac{k_\mu}{m_h} \tan(k_\mu R_{a\mu}) 
	 - Z \left( \frac{1}{m_e} + \frac{1}{m_h} \right) \frac{r_{1\mu} - R_{a\mu}}{r_{1a}}
\right]^2.
\end{multline}

\subsubsection{Confined trion}

 The wavefunction of X$^*$ reads:\cite{MovillaPRB} 
\begin{multline}
 \Psi_{X^*}(\mathbf{r}_1, \mathbf{r}_2,\mathbf{R}_a)  =  \\
	  \psi_{\text{box}}(\mathbf{r}_1)\, 
	  \psi_{\text{box}}(\mathbf{r}_2)\, 
	  \psi_{\text{box}}(\mathbf{R}_a)\, 
  J(r_1,r_2,r_{12}).
\end{multline}
\noindent Here, \(J(r_1,r_2,r_{12})\) is the Jastrow correlation factor:
\begin{equation}
\label{EqCT1}
J(r_1,r_2,r_{12}) = e^{-Z s/2} \, \cosh\left(\frac{Z Q t}{2}\right) \, 
e^{\frac{Z \beta r_{12}}{1+Z \alpha r_{12}}}
\end{equation}
\noindent with $Z$, $Q$, $\alpha$ and $\beta$ being variational parameters, 
while $r_1 = |\mathbf{r}_{e1} - \mathbf{r}_h|$, $r_2 = |\mathbf{r}_{e2} - \mathbf{r}_h|$, 
$r_{12} = |\mathbf{r}_{e1} - \mathbf{r}_{e2}|$, $s = r_1 + r_2$ and  $t = r_1 - r_2$.
 The local observable is:
\begin{equation}
\label{EqCT4}
O_N(\Psi_{X^*}(\mathbf{X})) = \left| \frac{1}{m_e} \nabla_1 \ln \Psi_{X^*} - \frac{1}{m_h} \nabla_h \ln \Psi_{X^*} \right|^2.
\end{equation}
\noindent And the final formula to be implemented:\cite{SuppMat}
\begin{multline}
\label{EqCT5}
O_N(\Psi_{X^*}(\mathbf{X})) =  \\
\sum_{\mu = x,y,z} \Bigg\{
\frac{1}{m_e} \left( -k_\mu \tan(k_\mu r_{e1,\mu}) \right)
- \frac{1}{m_h} \left( -k_\mu \tan(k_\mu r_{h,\mu}) \right) \\
 + \frac{1}{m_e} \left(
- \frac{Z}{2} + \frac{ZQ}{2} \tanh\left( \frac{ZQ t}{2} \right)
\right) \frac{r_{e1,\mu} - r_{h,\mu}}{r_1} \\
 + \frac{1}{m_e}
\left( \frac{Z \beta}{(1 + Z \alpha r_{12})^2} \right) 
\frac{r_{e1,\mu} - r_{e2,\mu}}{r_{12}} \\
 - \frac{1}{m_h} \left(
- \frac{Z}{2} - \frac{ZQ}{2} \tanh\left( \frac{ZQ t}{2} \right)
\right) \frac{r_{e2,\mu} - r_{h,\mu}}{r_2} \\
 - \frac{1}{m_h}
\left( \frac{Z \beta}{(1 + Z \alpha r_{12})^2} \right)
\frac{r_{e2,\mu} - r_{e1,\mu}}{r_{12}}
\Bigg\}^2.
\end{multline}

\subsubsection{Confined biexciton}

The wavefunction has the form:\cite{MovillaPRB}
%
\begin{multline}
	\Psi_{BX} =  \psi_{\text{box}}(\mathbf{r}_1)\, 
	  \psi_{\text{box}}(\mathbf{r}_2)\, 
	  \psi_{\text{box}}(\mathbf{R}_a)\, 
	  \psi_{\text{box}}(\mathbf{R}_b)\, \\ 
	\times F(r_{1a}, r_{1b}, r_{2a}, r_{2b}),
\end{multline}
\noindent where
\begin{equation}
\label{EqCBE2}
F = e^{-Z \frac{s_1 + s_2}{2}} \cosh\left[Z Q \frac{t_1 - t_2}{2}\right] e^{Z \frac{\beta r_{12}}{1 + Z \alpha r_{12}}} e^{Z \frac{\beta r_{ab}}{1 + Z \alpha r_{ab}}},
\end{equation}
\noindent with $Z$, $Q$, $\alpha$ and $\beta$ being variational parameters, 
$s_1 = r_{1a} + r_{1b}$, $s_2 = r_{2a} + r_{2b}$, $t_1 = r_{1a} - r_{1b}$, and $t_2 = r_{2a} - r_{2b}$.

\noindent The resulting expression for the local observable is:\cite{SuppMat}

\begin{multline}
\label{eq:BX}
O_N(\Psi_{BX}(\mathbf{X})) = \sum_{\mu=1}^{3} \bigg\{ \frac{1}{m_e} \bigg[
  - k_\mu \tan(k_\mu r_{1\mu})   \\
  - \frac{Z}{2} \left( \frac{r_{1\mu} - R_{a\mu}}{r_{1a}} + \frac{r_{1\mu} - R_{b\mu}}{r_{1b}} \right) \\
 + \frac{ZQ}{2} \tanh\left( \frac{ZQ}{2}(t_1 - t_2) \right)
           \left( \frac{r_{1\mu} - R_{a\mu}}{r_{1a}} - \frac{r_{1\mu} - R_{b\mu}}{r_{1b}} \right) \\
 + Z \beta \frac{r_{1\mu} - r_{2\mu}}{r_{12}(1 + Z \alpha r_{12})^2}
\bigg] \\
 - \frac{1}{m_h} \bigg[
  - k_\mu \tan(k_\mu R_{a\mu}) 
  - \frac{Z}{2} \left( \frac{R_{a\mu} - r_{1\mu}}{r_{1a}} + \frac{R_{a\mu} - r_{2\mu}}{r_{2a}} \right) \\
 + \frac{ZQ}{2} \tanh\left( \frac{ZQ}{2}(t_1 - t_2) \right)
             \left( \frac{R_{a\mu} - r_{1\mu}}{r_{1a}} - \frac{R_{a\mu} - r_{2\mu}}{r_{2a}} \right) \\
 + Z \beta \frac{R_{a\mu} - R_{b\mu}}{r_{ab}(1 + Z \alpha r_{ab})^2}
\bigg] \bigg\}^2
\end{multline}

\section{Results}

As an illustrative example, we calculate the characteristic frequency describing the internal motion of effective excitons, 
$\omega_A$, in cubic CsPbBr$_3$ NCs populated with X, X$^*$ or BX.  
The variational wavefunctions, $\Psi_X$, $\Psi_{X^*}$ and $\Psi_{BX}$, are obtained using the effective mass Hamiltonians described in Ref.~\cite{MovillaPRB},
which consider the influence of quantum confinement, dielectric confinement, polaronic interactions (with a Haken-like model) and electronic correlations.
Material parameters can be found in the same work. 
A few of them are summarized in Table \ref{table1}, since they can be relevant for the analysis. 
Note that we use the same polaron radius ($l$), static ($\varepsilon_0$) and dynamic ($\varepsilon_\infty$) dielectric constants for all excitonic complexes.

\begin{table}[h]
	\caption{A few relevant material parameters of CsPbBr$_3$: polaron radius, static and dynamic dielectric constant, outer dielectric constant, bulk LO phonon energy.\cite{MovillaPRB}}
\label{table1}
\begin{center}
\begin{ruledtabular}
\begin{tabular}{c  c  c  c  c}
	$l $ 	   & $\varepsilon_{0}$ & $\varepsilon_{\infty}$ & $\varepsilon_{out}$ & $\omega_{LO}$ \\ \hline
	 3.0 nm    &  16 	       & 4.5 		        & 2.56 		      & 18 meV\\
\end{tabular}
\end{ruledtabular}

\end{center}
\end{table}

\begin{figure}[h]
    \centering
    \includegraphics[width=7cm]{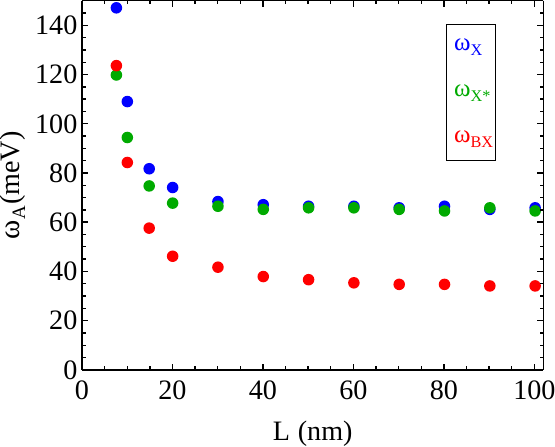}
    \caption{
	Frequency of the internal electron-hole motion for effective excitons inside X, X$^*$, and BX (blue, green, and red dots, respectively), 
	as a function of the lateral size of CsPbBr$_3$ NCs in the intermediate and weak confinement regime.
	$\omega_{BX}$ is systematically smaller than $\omega_{X}$. 
	$\omega_{X^*}$ matches $\omega_{X}$ in the weak confinement regime. 
}
    \label{fig1}
\end{figure}

Figure \ref{fig1} shows the frequencies of effective excitons in NCs with lateral sizes from $L=7$ nm to $L=100$ nm. 
This covers from intermediate to very weak (bulk-like) confinement regimes.\cite{ButkusCM}
The results indicate that the characteristic frequencies of all species decrease with the NC size.
The slower internal motion is due to the decrease of kinetic energy.
But the remarkable result here is that the frequency inside a BX ($\omega_{BX}$, red dots), remains significantly lower 
than that inside X ($\omega_X$, blue dots) regardless of $L$.
By contrast, the frequency inside $X^*$ ($\omega_{X^*}$, green dots) is smaller than $\omega_X$ in small NCs, but matches it in larger ones.
The same qualitative trends are observed if the distance dependence of the dielectric constant we used here is replaced by a homogeneous, 
effective dielectric constant (not shown), which means that it is electronic correlations --rather than subtle polaronic effects-- which 
determine the ratio of $\omega_X$, $\omega_{X^*}$ and $\omega_{BX}$.

The above results are in line with previous studies regarding the internal structure of weakly confined X$^*$ and BX in lead halide perovskite nanostructures. 
On the one hand, X$^*$ adopt a spatial configuration which can be sketched as a weakly bound extra charge orbiting around a nearly unperturbed exciton.\cite{ClimenteJPCc} 
 A similar frequency is then expected for the effective exciton as compared to that of an isolated X, $\omega_{X^*} \approx \omega_X$.
 On the other hand, BX have a more complex structure, 
consisting of two weakly interacting excitons that accommodate each other by slightly perturbing both subsystems.
One such perturbation is the increase of the intra-X electron-hole mean distance,\cite{ClimenteJPCL}
 which anticipates the slower internal motion and characteristic frequency observed in Fig.~\ref{fig1}, $\omega_{BX} < \omega_X$.
The different behavior for small $L$, where $\omega_{X^*}$ approaches $\omega_{BX}$,
is because the constraint of the particles into a smaller volume causes the extra charge in X$^*$ to become dynamically correlated with the electron-hole pair. 
The system behaves more like a compact object, yielding a more interdependent motion.\\ 

The internal dynamics described by Fig.~\ref{fig1} can explain the different dielectric constant sensed by X, X$^*$ 
and BX in weakly confined NCs.\cite{ChoACS,ZhuAM,MovillaPRB}
To this end, we translate internal frequences into an effective (static) screening picture 
by means of a two-step Lorentz-Drude model.\cite{Yu2001_book, Fox2001_book}  
 The real part of the field frequency-dependent dielectric constant, 
 which describes the dispersive behavior of the material, is given by:
\begin{multline}
\label{Lorentz-Drude}
\varepsilon_R(\omega) = \varepsilon_\infty + \frac{(\varepsilon_0 - \varepsilon_i)\, \omega_{{\rm ph}_1}^2 (\omega_{{\rm ph}_1}^2 - \omega^2)}{(\omega_{{\rm ph}_1}^2 - \omega^2)^2 + \gamma_1^2 \omega^2} f_1 \\
	+ \frac{(\varepsilon_i - \varepsilon_\infty)\, \omega_{{\rm ph}_2}^2 (\omega_{{\rm ph}_2}^2 - \omega^2)}{(\omega_{{\rm ph}_2}^2 - \omega^2)^2 + \gamma_2^2 \omega^2} f_2.
\end{multline}
\noindent Here, \( \varepsilon_0 \), \( \varepsilon_i \), and \( \varepsilon_\infty \) are the static, intermediate, and high-frequency dielectric constants, respectively; 
\( \omega_{{\rm ph}_j} \) are the characteristic (ionic and electronic) frequencies, \( \gamma_j \) the damping terms, and \( f_j \) weighting factors that determine the strength of each contribution.
Because we aim at a qualitative explanation, we minimize the adjustable parameters of the model by 
setting equal contribution strengths for the ionic and electronic contributions, $f_1 = f_2 = 1$,
 and $\varepsilon_i=0.45 \,(\varepsilon_0+\varepsilon_\infty)$.
The damping terms are set to $\gamma_1 = \gamma_2 = 4$ meV, 
a reasonable value about one order of magnitude smaller than the bulk longitudinal optical phonon mode, $\omega_{LO}$.
The characteristic frequencies are $\omega_{{\rm ph}_1}=\omega_{LO}$ and $\omega_{{\rm ph}_2}=2250$ meV.\cite{ManninoPCC} \\

\begin{figure}[h!]
    \centering
    \includegraphics[width=7cm]{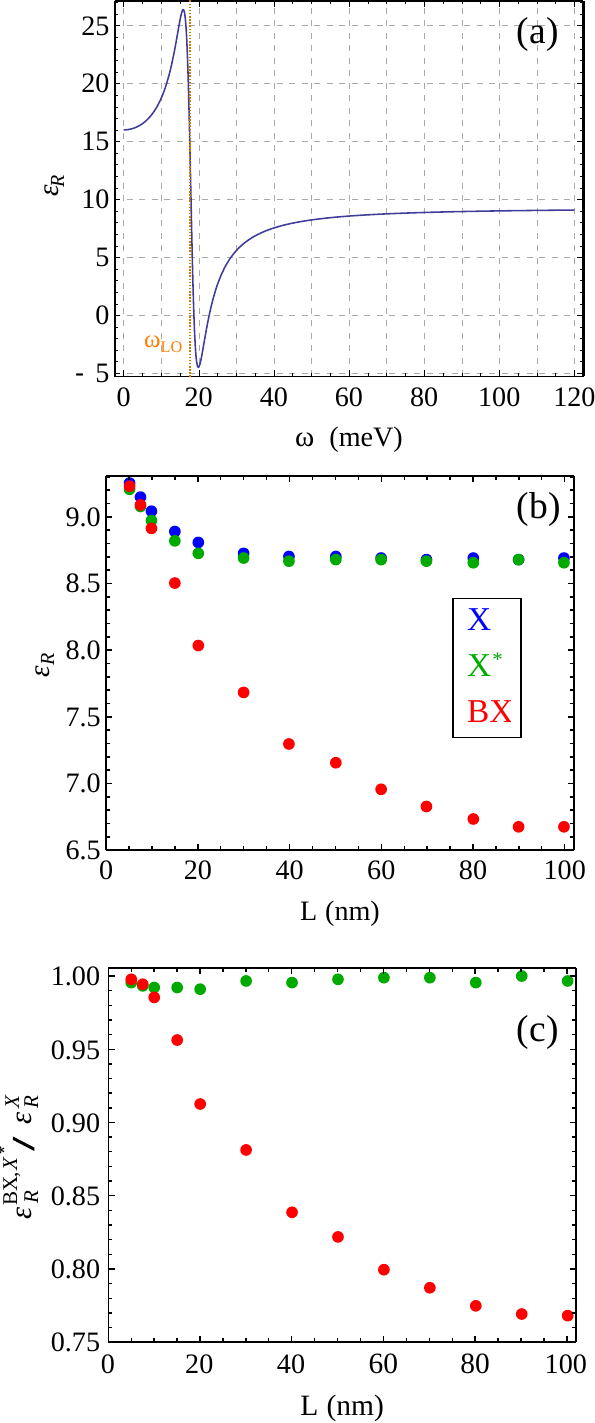}
    \caption{
	(a) Real part of the frequency-dependent dielectric function (Eq. (\ref{Lorentz-Drude})) in the vicinity of the longitudinal optical phonon resonance
	(represented by the dotted, vertical line).
	(b) Size dependence of the effective dielectric screening of X, X$^*$, and BX (blue, green, and red dots).
	The screening is calculated with the characteristic frequencies of Fig.~\ref{fig1}.
	(c) Dielectric screening experienced by X$^*$ and BX relative to that of X. 
	In the strong confinement regime, $\varepsilon_R^{X} \approx \varepsilon_R^{X^*} \approx \varepsilon_R^{BX}$.
	In the weak confinement one, $\varepsilon_R^{X} \approx \varepsilon_R^{X^*} > \varepsilon_R^{BX}$.
}
    \label{fig2}
\end{figure}

Figure \ref{fig2}a shows the profile of the ensuing frequency-dependent dielectric response $\varepsilon_{R}$ in the vicinity of the bulk LO phonon resonance. 
 One can see how the dielectric response decreases from \( \varepsilon_0 \) to \( \varepsilon_i \) as the frequency increases, 
 which reflects the deactivation of ionic polarization for rapidly oscillating fields. 
It should be noted, however, that all the excitonic frequencies in Fig. \ref{fig1} exceed $\omega_{LO}$, 
and therefore lie in a region where the dielectric screening increases with frequency.
 Consequently, $\omega_{BX} < \omega_{X^*} \approx \omega_X $ leads to $\varepsilon_R^{BX} < \varepsilon_R^{X^*} \approx \varepsilon_R^X$. 
 This is confirmed by Figure \ref{fig2}b, which shows the calculated dielectric constants for each excitonic complex throughout the entire $L$ range.\\
 
 Plotting $\varepsilon_R^{X^*}$ and $\varepsilon_R^{BX}$ relative to $\varepsilon_R^{X}$ (Fig. \ref{fig2}c) 
 allows us to make a few relevant observations.
 First off, for $L \lesssim 10$ nm, $\varepsilon_R^{X} \approx \varepsilon_R^{X^*} \approx \varepsilon_R^{BX}$. 
 This is because all the excitonic frequencies are well above that of $\omega_{LO}$,
 where the lattice ionic response is deactivated --high frequencies plateau in Fig.~\ref{fig2}a--.
 This result is in agreement with the literature of usual, strongly confined colloidal quantum dots, 
 where the physics of charged and multiexciton states has been traditionally understood by taking 
 the same  constant for all species.
 Second, as we deepen into the weak confinement regime ($L>10$ nm), the BX experiences reduced screening.
 This is because now excitonic frequencies lie closer to $\omega_{LO}$, where the lattice response becomes more sensitive to changes.
 For NC sizes around $15-20$ nm, $\varepsilon_{R}^{BX}$ reaches a $5-10\%$ decrease, 
 which is actually the value postulated by previous works to reconcile $\Delta_{BX}$ calculations with 
 experimental measurements in similar sizes.\cite{MovillaPRB, ZhuAM,  ChoACS} 
 Third, no substantial reduction in $\varepsilon_{R}^{X^*}$ is observed in this regime. 
 This is because $\omega_{X^*} \rightarrow \omega_X$.
 The fact that $\varepsilon_R^{X^*} \approx \varepsilon_R^X$ is at odds with speculative discussions in Refs.~\cite{ChoACS,NguyenPRB}.
 which invoked a reduced screening not only for BX, but also for X$^*$, for their computational model to reproduce experimental values of $\Delta_{X^*}$.
 It should be noted, however, that more sophisticated theoretical models, including dielectric confinement through $\varepsilon_{out}$,
 did not need to make such an assumption.\cite{ZhuAM,MovillaPRB}\\

\section{Conclusions}

We have developed a theoretical framework to calculate the characteristic frequency $\omega_A$ 
associated with the relative motion between one electron-hole pair inside an excitonic complex, 
under the influence of the remaining particles.
The frequency is expressed as the ratio of the square root of the expectation value 
of the squared commutator $[H, A]^2$ to the expectation value of the observable 
$A^2 = r_{eh}^2$, where $r_{eh}$ is the electron-hole distance. 
Building on variational Quantum Monte Carlo wavefunctions that incorporate confinement and correlation effects,
explicit expressions have been derived for X, X$^*$ and BX confined in cuboidal NCs.

When applied to CsPbBr$_3$ NCs, our model shows that BX have lower characteristic frequencies than X,
 as inter-X interactions slow down the intra-X motion. 
 By contrast, the frequency in X$^*$ approaches that of X with increasing NC size,
 which is a consequence of the extra carrier exerting a weak perturbation on the excitonic motion.\\

 These characteristic frequencies have direct implications on the lattice screening.
 In the strong confinement regime, $\omega_A \gg \omega_{LO}$, so that ionic screening is almost completely deactivated.
 In the weak confinement one, $\omega_A \gtrsim \omega_{LO}$, which makes the lattice sensitive to the differences between X, X$^*$ and BX systems.
 Specifically, we show that $\varepsilon_R^{BX} < \varepsilon_R^{X^*} \approx \varepsilon_R^{X}$ 
 (an apparent paradox, because $\omega_{BX} < \omega_X$).
 This result is consistent with the reported need for a reduced effective dielectric constant 
 in static BX binding energy calculations. 
 Our approach provides a dynamic and physically motivated framework for understanding excitonic interactions in confined systems.

\begin{acknowledgments}
We acknowledge support from Grant No. PID2021-128659NB-I00, funded by Ministerio de Ciencia e Innovaci\'{o}n (MCIN/AEI/10.13039/501100011033 and ERDF ``A way of making Europe'').
\end{acknowledgments}

\bibliographystyle{apsrev4-2}
\bibliography{pv_nc}

\end{document}